\documentclass[11pt]{article}
\usepackage{amsmath}
\usepackage{amssymb}
\usepackage{graphicx}
\usepackage{diagbox}
\usepackage{hyperref}
\usepackage{multirow}

\renewcommand{\baselinestretch}{1.3}
  \renewcommand{\arraystretch}{1.1}
\begin{document}

 \title{Analysis of Nederlof's Algorithm for Subset Sum}

  \author{Zhengjun Cao$^{1}$,\quad Zhen Chen$^{1}$, \quad Lihua Liu$^{2,*}$}
  \footnotetext{$^{1}$Department of Mathematics, Shanghai University, Shanghai,  China. \\
    $^2$Department of Mathematics, Shanghai Maritime University, Shanghai,  China. \,  $^*$\,\textsf{liulh@shmtu.edu.cn}}

 \date{}\maketitle

\begin{quotation}
 \noindent\textbf{Abstract}.
 We show that  Nederlof's algorithm [\textit{Information Processing Letters}, 118 (2017), 15-16]  for constructing a proof that the number of subsets  summing to a particular integer equals a claimed quantity is flawed because: 1) its consistence is not kept; 2) the proposed recurrence formula is incorrect.


  \end{quotation}

\section{Introduction}

In computer science the subset sum problem is that: given a set (or multiset) of integers, is there a non-empty subset whose sum is equal to a given integer? In 1955, Gupta ever proved that \cite{G55}:

\textbf{Theorem 1}  \textit{Let $P(A, t)$ denote the number of partitions of $t$ into members of the set $A=\{a_0, a_1, a_2, \cdots, a_n\}$
($a_i$ distinct positive integers), $a_0=1$. Then
$${t+n \choose n} \leq P(A, t)\prod_{j=1}^n a_j\leq {t+\sum_{j=1}^n a_j \choose n} $$}

In 1956, Bateman and Erd\H{o}s proved that \cite{BE56}

\textbf{Theorem 2} \textit{If $A$ is any non-empty set of positive integers, then the number of partitions of $n$ into members of the set $A$, $P(A, n)$,  is a non-decreasing function of $n$ for large $n$, if and only if $A$ either:
(i) contains the element $1$ or (ii) $A$ contains more than one element and, if we remove any single element
from $A$, the remaining elements have greatest common divisor $1$.}

In complexity theory, a proof system for subset sum problem is referred to as a Merlin-Arthur protocol.
 Babai and Moran \cite{BM88} ever discussed the Arthur-Merlin games and a hierarchy of complexity classes.  In 2016, Williams \cite{W16} pointed out
the relation between strong ETH and Merlin-Arthur proof system.  Austrin et al. \cite{A16} pointed out that a special case of sub-set sum may be the hardest.  In 2017, Nederlof \cite{N17} proposed an algorithm for subset sum problem. In this note, we show that Nederlof algorithm is flawed.

 \section{Review of Nederlof's algorithm}

 The  subset sum problem discussed by Nederlof  \cite{N17} is that:
    given positive integers $w_1, \cdots, w_n$ along with a target integer $t$,  the task is to determine whether there exists a subset $X\subset \{1, \dots, n\}$ such that
    $$w(X) :=\sum_{i\in X} w_i=t. \eqno(1)$$
     Such an $X$ is referred to as a solution.
 The Nederlof's algorithm aims to construct a proof that the number of solutions of $(w_1, \cdots, w_n; t)$ is $c_t$.
  To do so, the prover and the verifier execute the following algorithms, respectively.

\noindent\rule[-0.25\baselineskip]{0.85\textwidth}{0.25mm}\vspace*{2mm}

{\renewcommand{\baselinestretch}{.95}
\renewcommand{\arraystretch}{.9}
 \parskip -1.0mm

Algorithm P$(w_1, \cdots, w_n; t)$.\quad Prove that the number of solutions is $c_t$.

Output: Prime $p=\Theta(\sqrt{nt}), c_i:|\{X\subset[n]: w(X)=i\}|$ for $i\leq nt: i\equiv_{p} t$.

\hspace*{3mm}1: \hspace*{4mm} Initiate $T[0, 0]=1$ and $T[0, i]=0$ for $0<i\leq nt$.

\hspace*{3mm}2: \hspace*{4mm}  \textbf{for} $j=1 \rightarrow n$ \textbf{do}

\hspace*{3mm}3: \hspace*{4mm} \hspace*{4mm}  \textbf{for} $i=1 \rightarrow nt$ \textbf{do}

\hspace*{3mm}4: \hspace*{4mm} \hspace*{4mm}\hspace*{4mm} \textbf{if} $i<w_j$ \textbf{then}

\hspace*{3mm}5: \hspace*{4mm}  \hspace*{4mm}\hspace*{4mm}\hspace*{4mm} $T[j, i]\leftarrow T[j-1, i]$

\hspace*{3mm}6: \hspace*{4mm} \hspace*{4mm}\hspace*{4mm} \textbf{else}

\hspace*{3mm}7: \hspace*{4mm} \hspace*{4mm}\hspace*{4mm}\hspace*{4mm} $T[j, i]\leftarrow T[j-1, i]+T[j-1, i-w_j]$

\hspace*{3mm}8: \hspace*{4mm} Pick the smallest prime $p$ such that $2\sqrt{nt}<p<4\sqrt{nt}$.

\hspace*{3mm}9: \hspace*{4mm} \textbf{for} $i\leq nt$ such that $i\equiv_{p}  t$ \textbf{do}

\hspace*{2.5mm}10: \hspace*{4mm} \hspace*{4mm} $c_i \leftarrow T[n, i]$.

\hspace*{2.5mm}11:\hspace*{4mm} \textbf{return} $(p, \{c_i\})$.\vspace*{1mm}

Algorithm V$(w_1, \cdots, w_n; t;  p, \{c_i\})$.\quad Verify the proof for number of solutions.

Output: $c_t$, if the proof is as output by P, NO with 1/2 probability otherwise.

\hspace*{3mm}12: \hspace*{4mm} Pick a prime $q$ satisfying $2^nt<q<2^{n+1}t$ and a random $r\in \mathbb{Z}_q$.

\hspace*{3mm}13:  \hspace*{4mm} Initiate $T'[0, 0]=1$ and $T'[0, i]=0$ for $0<i<p$.

\hspace*{3mm}14:  \hspace*{4mm} \textbf{for} $j=1\rightarrow n$ \textbf{do}

\hspace*{3mm}15:  \hspace*{4mm}\hspace*{4mm}  \textbf{for} $i=1\rightarrow p$ \textbf{do}

\hspace*{3mm}16:   \hspace*{4mm}\hspace*{4mm}\hspace*{4mm} $T'[j, i]\leftarrow (T'[j-1, i]+r^{w_j}\cdot T'[j-1, (i-w_j) \,\%\, p])\,\%\, q$.

\hspace*{4mm}\hspace*{4mm}\hspace*{4mm}\hspace*{4mm}\hspace*{6mm} $x\%\,p$ denotes remainder of $x$ divided by $p$

\hspace*{3mm}17: \hspace*{4mm} Compute $\sum_i c_i r^i\%\,q $.

\hspace*{3mm}18: \hspace*{4mm} \textbf{if} $\sum_i c_i r^i \equiv_q T'[n, t\,\%\,p]$ \textbf{then return} $c_t$ \textbf{else return} NO. }

\noindent\rule[0.55\baselineskip]{0.85\textwidth}{0.25mm}

\section{Analysis of Nederlof's algorithm}\parskip 1mm


\subsection{Inconsistency}

The correctness of Nederlof's algorithm was not explained explicitly. For example, the choice of the prime $p$ and
the correctness of the recurrence formula
$$T'[j, i]= (T'[j-1, i]+r^{w_j}\cdot T'[j-1, (i-w_j) \,\%\, p])\,\%\, q   $$
 are not explained. Besides, the initial values $T[1, 0], T[2, 0], \cdots, T'[1, 0], T'[2, 0], \cdots$, are not specified. We find its consistency is not kept.
  To see this flaw, it suffices to investigate the following example.

\textbf{Example 1}.   Suppose that $w_1=1, w_2=2, w_3=3, w_4=4; t=17$. Then $ n=4, nt=68, c_{17}=0, 2\sqrt{nt}=2\sqrt{68}\thickapprox 16.492$ and $p=17$. By Nederlof's algorithm, we have
$$ T[j, i]=\left\{
\begin{array}{l l}
   T[j-1, i], & i<w_j  \\
 T[j-1, i]+T[j-1, i-w_j], &   i\geq w_j
\end{array}
\right. \eqno(2)$$
for $i=1, \cdots, 68; j=1, 2, 3, 4$. Hence,
\begin{align*}
T[0, 0]=&1, T[1, 1]=T[0, 1]+T[0, 0]=1,\\
  T[2, 2]=&T[1, 2]+T[1, 0]=T[0, 2]+T[0, 1]+T[1, 0]=T[1, 0], \cdots,\\
 c_{17}=&T[4, 17]=T[3, 17]+T[3, 13]=\cdots=0,\,\,  c_{34}= c_{51}= c_{68}=0,
\end{align*}
  The prover's output is $ p=17,  c_{17}=0, c_{34}=0, c_{51}=0,  c_{68}=0$.

Suppose that the verifier picks $q=277, r=7$. Then $\sum_i c_i r^i =7$.

By the recurrence formula
$$T'[j, i]=(T'[j-1, i]+r^{w_j}\cdot T'[j-1, (i-w_j) \,\%\, p])\,\%\,q \eqno(3)$$
we have
\begin{align*}
T'[n, t\,\%\,p]&=T'[4, 20\% 19]=T'[4, 1]\\
&=(T'[3, 1]+7^4 T'[3, (1-4)\,\%\,19])\,\%\,331=(T'[3, 1]+7^4 T'[3, 16])\,\%\,331\\
&=(T'[2, 1]+7^3T'[2, 17]+7^4 T'[2, 16]+7^7T'[2, 13])\,\%\,331\\
&=\cdots =7\,T'[0, 0]\,\%\,331.
\end{align*}
To ensure $\sum_i c_i r^i \equiv_q T'[n, t\,\%\,p]$, \underline{it has to specify $T'[0, 0]=0$}. This leads to a contradiction. That means the consistence of the algorithm is not kept.

\subsection{The correct recurrence formula}

   Suppose that $A=\{w_1, w_2, \cdots, w_n\}$, all members are positive integers and $w_i\not=w_j$ for $i\not = j$. Let $A_j=\{w_1, w_2, \cdots, w_j\}$, $1\leq j\leq n$.
  Denote the number of partitions of $i$ into members of the set $A_j$ by $T[A_j, i]$.
   Then the following recurrence formula
$$ T[A_j, i]=\left\{
\begin{array}{l l}
   T[A_{j-1}, i], & i<w_j  \\
 T[A_{j-1}, i]+T[A_{j-1}, i-w_j], &   i\geq w_j
\end{array}
\right. \eqno(4)$$
holds, where $T[A_{j-1}, i]$ is the number of solutions which do not contain $w_j$ and $T[A_{j-1}, i-w_j]$ is the number of solutions which do contain $w_j$ exactly once. Apparently, Nederlof \cite{N17} confused Eq.(2) with Eq.(4).

\section{Conclusion} We analyze the Nederlof's algorithm for constructing a proof that the number of subsets summing to an integer is a claimed quantity. We also remark that it is somewhat difficult to theoretically compute the number of partitions of an positive integer into members of a finite set.


\end{document}